\documentstyle[aps,preprint,epsf]{revtex}

\begin{document}

\draft

\title{Isospin symmetry breaking nucleon-nucleon potentials \\ 
       and nuclear structure}
 
\author{H.\ M\"{u}ther} 
\address{Institut f\"ur Theoretische Physik, Universit\"at T\"ubingen,
         D-72076 T\"ubingen, Germany}
\author{A.\ Polls}
\address{Departament d'Estructura i Costituents de la Mat\`eria,
         Universitat de Barcelona, E-08028 Barcelona, Spain}
\author{R.\ Machleidt} 
\address{Department of Physics, University of Idaho, Moscow, 
         ID 83844, U.S.A.}

\maketitle

\begin{abstract}

Modern nucleon-nucleon (NN) potentials, which accurately fit the nucleon-nucleon
scattering phase shifts, contain terms which break isospin symmetry. 
The effects of these symmetry violating terms on the bulk properties
of nuclear matter are investigated. The predictions of the charge symmetry
breaking (CSB) terms are compared with the Nolen-Schiffer (NS) 
anomaly regarding the energies of neighboring mirror nuclei. 
We find that, for a quantitative explanation of the NS anomaly,
it is crucial to include CSB in partial waves with $L>0$
(besides $^1S_0$) as derived from a microscopic model for
CSB of the NN interaction.

\end{abstract}


{\it Isospin symmetry} (or charge independence) is invariance under any 
rotation in isospin space. 
Due to the mass difference between up and down quarks and due to the
electromagnetic interaction, this symmetry is slightly violated
which is referred to as isospin symmetry breaking (ISB)
or charge independence breaking.
{\it Charge symmetry} is invariance under a rotation by 180$^0$ about the
$y$-axis in isospin space if the positive $z$-direction is associated
with the positive charge.
The violation of this symmetry is known as charge symmetry breaking (CSB).
Obviously, CSB is a special case of ISB.

ISB of the strong nucleon-nucleon (NN) interaction means that,
in the isospin $T=1$ state, the
proton-proton ($T_z=+1$), 
neutron-proton ($T_z=0$),
or neutron-neutron ($T_z=-1$)
interactions are (slightly) different,
after electromagnetic effects have been removed.
CSB of the NN interaction refers to a difference
between proton-proton (pp) and neutron-neutron (nn)
interactions, only. 
For reviews on these matters, see Refs.~\cite{MO95,miller}.

In recent years, a new generation of realistic NN potentials 
has been developed which take ISB of the NN interaction into account
and, therefore, yield very accurate fits of the proton-proton
and proton-neutron (pn) scattering phase shifts\cite{nim,v18,cdbonn}. Since
these fits are based on the same phase shift analysis by the Nijmegen
group\cite{nimpsa} and yield a value for the $\chi^2$/datum 
very close to one, these various
potentials could be called phase-shift equivalent NN interactions. The
interactions by the Nijmegen group (Nijm1, Nijm2, Reid93)~\cite{nim}, 
the Argonne group (ArgV18)~\cite{v18} and the charge-dependent 
Bonn potential (CDBonn)~\cite{cdbonn} describe the
long-range part of the NN interaction in terms of the one-pion-exchange model,
accounting for the mass-difference between $\pi^0$ and the charged pions $\pi^+$
and $\pi^-$. This distinction between the exchange of a neutral pion and charged
pions is one origin of ISB in the resulting NN
interactions. It yields different phase shifts for $pp$ scattering (only
neutral pion exchange) as compared to $pn$ scattering in 
partial waves with isospin $T=1$. Additional ISB 
terms have to be included in the $^1S_0$ state to achieve
an accurate fit of $pp$ and $pn$ phase shifts in that partial wave. 

The new NN interactions also account for the mass difference 
between proton and neutron. 
This gives rise to a difference in the matrix elements of the  
meson-exchange interaction between two protons and two neutrons. This
breaks charge-symmetry, but only by a very small amount.
Additional CSB terms are included in the ArgV18 and CDBonn
potentials to reproduce the difference in the empirical $nn$ and $pp$ 
scattering lengths, after subtracting the effects of 
electromagnetic interactions. 

It is the aim of this brief report to study the effects of ISB 
on the calculated bulk properties of nuclear systems. 
Are these symmetry breaking
effects similar in all these NN interactions? What is the influence of the ISB
terms on the calculated symmetry energy and saturation properties of
nuclear matter? And more specifically, do the CSB terms 
explain the so-called Nolen-Schiffer
anomaly\cite{nolen}, the energy difference between neighboring mirror nuclei,
which cannot be explained by the electromagnetic interaction?
 
For that purpose we have performed Brueckner-Hartree-Fock (BHF) calculations of
nuclear matter, in which we distinguish between pp, pn and nn interactions. As a
reference we have also performed BHF calculations, in which the pp and nn
interactions have been replaced by the corresponding pn interaction. The
Bethe-Goldstone equation has been solved by determining a self-consistent
single-particle spectrum for the hole states, which is extended in a continous
way to states with momenta $k$ larger than the Fermi momentum $k_F$. 
For very high momenta, for which this prescription would yield a repulsive 
single-particle potential, the single-particle energy has been assumed to be 
identical to the kinetic energy. Such a continous choice for the 
single-particle spectrum yields larger binding energies than the conventional 
choice (single-particle energy equal to kinetic energy for $k \ge k_F$) and
seems to reduce the effects of three-body correlations in the hole-line
expansion\cite{baldo}.

The interactions CDBonn, ArgV18, Nijm1, Nijm2 and Reid93   
are quite different, although they are phase-shift equivalent and agree to a 
large extent in the OPE contribution. This is demonstrated in Table
\ref{tab:one}, which presents some results calculated for nuclear matter at the
empirical value for the saturation density ($\rho_0$ = 0.17 fm$^{-3}$, which
corresponds to a Fermi momentum of symmetric nuclear matter of $k_F$ = 1.36
fm$^{-1}$). The results for the energy of nuclear matter calculated in the 
Hartree-Fock (HF) approximation ($E_{HF}$, second column of Table
\ref{tab:one}) range from 6.06 MeV per nucleon obtained from the CDBonn
potential to 35.62 MeV per nucleon in the case of the Nijm2 interaction. The
discrepancy between these HF results and the empirical value of --16 MeV per
nucleon reflects the importance of NN correlations to be included in the nuclear
structure calculation. It is worth noting that these modern models of the NN
interaction are much ``softer'' than older versions of a realistic NN
interaction: e.g.~the Reid soft-core potential\cite{reid} yields a HF energy of 
175 MeV per nucleon. It should also be mentioned that the NN interactions, which
contain non-local terms like the Nijm1 and the CDBonn potential are softer,
i.e.~yield less repulsive HF energies, than the NN interactions defined in terms
 of local potentials\cite{morten}.

If effects of NN correlations are included by employing the BHF approach, the
various models of the NN interaction yield results (see first column in Table
\ref{tab:one}) which are rather close to each other, the largest difference
being 3 MeV, and also fairly close to the empirical energy. The NN interactions
with weaker tensor force (CDBonn and ArgV18)\cite{deuter} yield 
more binding energy than those which contain a stronger tensor force and the
softer nonlocal potentials CDBonn and Nijm1 tend to predict larger binding
energies than corresponding local interactions. This can easily be understood:
Stiff potentials and those with a large tensor component, receive a large part
of the attraction in the T-matrix from terms of second and higher order in the
potential $V$. Due to Pauli and dispersion effects these attractive
contributions are quenched in the $G$-matrix. If two NN interactions fit the
same NN phase shifts, therefore yield the same T-matrix, the $G$-matrixelements
will be less attractive for a stiff potential as compared to a soft one.

\begin{figure}[t]
\epsfysize=9.0cm
\begin{center}
\makebox[16.4cm][c]{\epsfbox{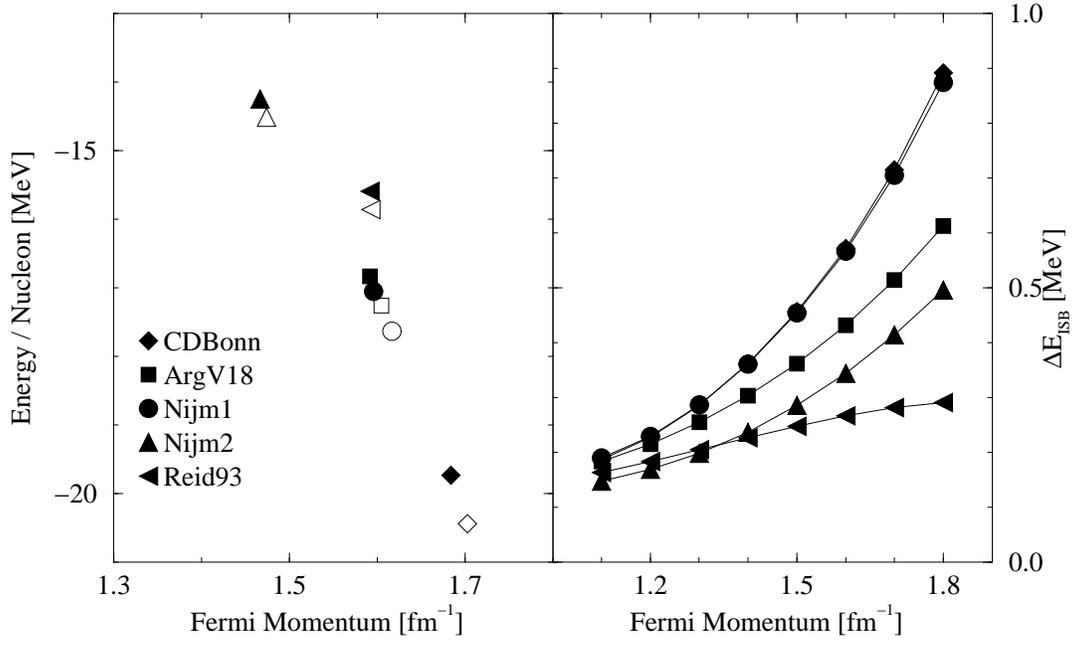}}
\end{center}
\caption{The left part of this figure displays the saturation points for
symmetric nuclear matter evaluated in the BHF approximation for various NN
interactions. Filled symbols represent the results for the interaction with
inclusion of ISB, while open symbols are obtained if the
$pn$ interaction is used for all two-nucleon pairs. 
The right part of the figure shows $\Delta E_{ISB}$
as defined in Eq.~(\protect{\ref{eq:one}}).}
\label{fig:one}
\end{figure}

The differences in the predictions for the binding energies are enhanced, if
we compare the saturation points, i.e.~the minima of the energy versus density
curves (left part of Fig.~\ref{fig:one}), 
instead of the energies calculated at the empirical saturation density.
Notice that the saturation points 
(solid symbols) determined from these modern potentials fall on the so-called
Coester band\cite{coester}.

The main interest of the present study is to explore the effects of
ISB in these modern NN interactions. For that purpose we
repeated the BHF calculations of nuclear matter replacing the $pp$ and $nn$
interactions by the corresponding matrix elements of the $pn$ interaction,
i.e., we assume perfect isospin symmetry and identify the NN interaction with
the $pn$ interaction. If we
denote the energy from this isospin symmetric calculation by $E_{BHF}^{np}$, 
then we may define an energy correction
which is due to the breaking of isospin-symmetry by
\begin{equation}
\Delta E_{ISB} = E_{BHF} - E_{BHF}^{np}\, . \label{eq:one}
\end{equation}
Results for this energy correction are displayed in Table \ref{tab:one} and
the right part of Fig.~\ref{fig:one}. 

The energy correction is repulsive
reflecting the fact that the $pn$ interaction is more attractive in the partial
waves with isospin $T=1$ than the corresponding $pp$ and $nn$ interactions. A
main contribution of this ISB can be related to the
pion-exchange contribution, which is repulsive in the dominant $^1S_0$ partial
wave. While the one-pion-exchange (OPE) contribution to the $pp$ and $nn$
interaction can only be mediated by the neutral pion $\pi^0$, charged pions
$\pi^{\pm}$ can be exchanged in the $pn$ interaction. Since the mass of the
$\pi^0$ is smaller than the mass of $\pi^{\pm}$, the OPE contribution
is more repulsive in $pp$ and $nn$ than in the $pn$ interaction\cite{li2}.
The ISB effect is largest and essentially identical 
for the CDBonn and Nijm1 potentials (the two curves refering to these potentials
can hardly be separated in the right part of Fig.~\ref{fig:one}). 
It is weaker by roughly 35 percent for the potentials Nijm2 and Reid93, while 
the prediction of the Argonne V18 potential is in between.

Since the Argonne V18 potential is local, one would expect that it predicts
$\Delta E_{ISB}$ very similar to the other local potentials,
Nijm2 and Reid93. However, while 
the latter two potentials both predict $\Delta E_{ISB}=0.22$ MeV,
the ArgV18 result is 0.28 MeV, at $k_F=1.36$ fm$^{-1}$ (cf.\ Table I).
The  reason for this  discrepancy is as follows.
Since the pp data are the most precise and reliable ones, all new
high-precision ISB potentials are produced by first
constructing the pp version of the NN potential with an acurate fit
to the pp data. The $T=1$ np potential is then
{\it defined} as the pp potential, but with the $\pi^0$ exchange
replaced by the $\pi^0/\pi^\pm$ exchanges appropriate
for $T=1$ np and with one proton mass replaced by a neutron
mass. The change in OPE explains about 50\% of the empirically
known difference between the pp scattering length (corrected for
electromagnetic effects) and the np one, in the $^1S_0$ state.
The remaining 50\% can---to a large extend, 
but not completely---be explained from ISB that emerges from
2$\pi$ and $\pi\alpha$ exchanges where $\alpha$ denotes
a heavy meson~\cite{MO95,li2}.
However, in the current high-precision NN potentials the latter
is not included as derived from a microscopic model~\cite{li2};
instead, what is missing to get the $^1S_0$ $np$ scattering length right,
is just fitted by enhancing a parameter in the model that
describes the intermediate range attraction.
Since the Nijmegen and the CDBonn potentials are constructed in
a partial wave basis, this fitting affects only the $^1S_0$
state, making this state more attractive for $np$. 
The procedure is different for ArgV18.
The Argonne V18 potential is constructed  in a $(S,T)$
decomposition (where $S$ denotes the total spin 
and $T$ the total isospin of the two-nucleon system).
For getting the $^1S_0$ $np$ scattering length correct, the 
$(S=0, T=1)$ 
potential is made slightly more attractive. However, in terms of
a partial wave decomposition, this implies that all partial wave
states with $(S=0, T=1)$, namely, $^1S_0$, $^1D_2$, $^1G_4$, etc.,
have their attraction enhanced. This inceases the binding energy
obtained from the Argonne np potential 
(as compared to $np$ potentials that adjust only
$^1S_0$) and, thus, leads to a larger $\Delta E_{ISB}$ 
of 0.06 MeV for ArgV18.

The discussion of the previous paragraph raises the question,
what change in $\Delta E_{ISB}$ is obtained if the `correct'
ISB in partial waves other than $^1S_0$ is applied
(instead of just extrapolating what fits the $np$ singlet
scattering length).
Unfortunately, we do not have any reliable empirical information
on the ISB of the nuclear force in partial waves
with $L>0$ (where $L$ denotes the total orbital angular momentum of the
two-nucleon system), at this time. However, there are microscopic models
that predict ISB. If those models predict the empirically known
ISB of the $^1S_0$ scattering about correctly, then one may imply
that the predictions for higher partial waves are also reasonable.
One such calculation is published in Ref.~\cite{li2},
which is based upon the Bonn full model for the 
NN interaction~\cite{MHE87}.
A refined version of the CDBonn potential~\cite{Mac98}
has been constructed
(that has become known as the ``CDBonn99'' potential~\cite{HF98})
which takes the ISB effects as predicted in Ref.~\cite{li2}
in partial waves with $L>0$ into account and, in addition,
includes the ISB effects from irreducible $\pi\gamma$ exchange as
derived in Ref.~\cite{Kol98}.
We have used this CDBonn99 model~\cite{Mac98} (in short: `CDBo99')
in our calculations as well
and obtain $\Delta E_{ISB}=0.370$ MeV at $k_F=1.36$ fm$^{-1}$ (see table 
\ref{tab:one}).
This is to be compared to 0.329 MeV as obtained from the ordinary
CDBonn~\cite{cdbonn}.
Thus, the refined treatment of ISB in partial waves higher
than $^1S_0$ increases $\Delta E_{ISB}$ by 0.04 MeV.

A comparison with the Argonne V18 result discussed above
implies that ArgV18 contains a reasonable estimate for the ISB effects 
from higher partial waves. ArgV18 
is overestimating the ISB effect from higher partial waves
by roughly 50\%. The reason for this is presumably that the ISB term in
the ArgV18 potential (that describes ISB beyond OPE
in a phenomenological way) is too long-ranged.

The effect of ISB is also reflected in the saturation
points displayed in the left part of Fig.~\ref{fig:one}. While the filled
symbols refer to the saturation points predicted from the calculation with
inclusion of the ISB terms, the open symbols are obtained
for the corresponding $E^{np}$. The effect of the ISB 
terms on the calculated symmetry energy (difference between binding energy per
nucleon of neutron matter and nuclear matter at the same density) is negligible
(see Table \ref{tab:one}).

\begin{figure}[h] 
\epsfysize=9.0cm
\begin{center}
\makebox[16.4cm][c]{\epsfbox{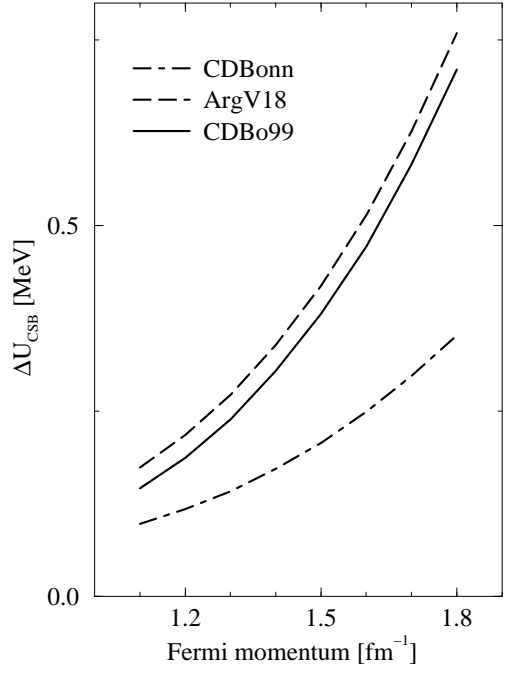}}
\end{center}
\caption{Difference between the single-particle potentials 
for protons and neutrons $\Delta U_{CSB}$ [as defined 
in Eq.~(\protect{\ref{eq:two}})] 
calculated at $k=k_F$ for various densities and three different NN potentials.}
\label{fig:two}
\end{figure}

The NN interactions CDBonn and ArgV18 have been adjusted to reproduce the
differences in the scattering length and effective range parameters for $pp$ and
$nn$ scattering. Therefore these two potentials as well as the NN interaction 
CDBonn99 also yield significant differences in 
predicting  the single-particle potentials for protons $U_p(k)$ and 
neutrons $U_n(k)$ as a function of the momentum $k$ in symmetric 
nuclear matter. It turns out that the momentum dependence of the difference
\begin{equation}
\Delta U_{CSB} (k) =  U_p(k) -  U_n(k) \, , \label{eq:two}
\end{equation}
is weak. Therefore we display in Fig.~\ref{fig:two} this difference just 
calculated at the Fermi momentum $k=k_F$ for various densities. This difference
is positive, which implies that the $pp$ interaction is less attractive than the
corresponding $nn$ interaction as one can already deduce from the difference in
the $^1S_0$ scattering lengths ($a_{pp}$=--17.3 fm compared to $a_{nn}$=--18.8
fm)\cite{li1}. 
  
At the empirical saturation density, the ArgV18 and CDBonn99 predict a value for
$\Delta U_{CSB}$ of 0.31 MeV and 0.28 MeV, respectively, while the CDBonn 
potential yields 0.16 MeV. These
numbers should be compared with a value of 0.2--0.3 MeV, which is typical for the
Nolen Schiffer anomaly, i.e.~the energy difference between the 
binding energies of
mirror nuclei, which is beyond the effects of the electromagnetic
interaction\cite{miller}.
It seems that the BHF predictions based on the three potentials yield the right
order of magnitude. 
  
At a first glance, it is disturbing that some predictions differ by
almost a factor of two.
However, this can be explained.
The reason is similar to what we discussed above in conjunction
with $\Delta E_{ISB}$.
In all models, the construction of the $nn$ potential starts
from the $pp$ potential. To fit the slightly more attractive
$^1S_0$ $nn$ scattering length, the attraction in the potential
is slightly enhanced.
In the case of the Argonne V18 potential, this implies more
attraction for all $(S=0, T=1)$ $nn$ partial waves. The CDBonn99 potential,
which contains the microscopically determined CSB effects,
predicts different interactions in all $T=1$ partial waves.
In the CDBonn, strictly only the $^1S_0$ state is made slightly
more attractive to match $a_{nn}$. This explains the larger value
for $\Delta U_{CSB}$ obtained for ArgV18 and CDBonn99. Obviously,
CSB in partial waves higher than $^1S_0$ is of considerable
influence on $\Delta U_{CSB}$.

Therefore our conclusion is the following:
to reproduce the Nolen Schiffer
anomaly, it is insufficient to take just the CSB in the $^1S_0$
scattering length into account; a distinguished knowledge of CSB
in partial waves beyond $^1S_0$ is crucial. 

%This can only be determined from 
%microscopic calculations and not from extrapolations as
%it is done in phenomenological models like ArgV18.

In summary, 
we have calculated ISB effects in nuclear matter and find
that they are generally small, but important in some cases.
For the binding energy per nucleon in symmetric nuclear
matter, the breaking of isospin-symmetry is a very small effect. For all the potentials
analyzed in the paper, ISB produces a small loss of binding energy 
(as compared to calculations that use the $np$ potential throughout),
which is mainly caused by
 the ISB in the $^1S_0$ partial wave. The effects in the symmetry energy are 
essentially
 negligible. On the other hand, if one wants to explain the Nolen Schiffer anomaly
 by calculating the difference between the single particle potential for protons and
 neutrons, which is a measure for charge symmetry breaking, one finds that,
in order to have quantitative agreement, it is necessary to include CSB in partial
waves beyond $^1S_0$. A natural way to incorporate CSB in higher partial waves, based on 
microscopic calculations, is provided by the recent update of the
CDBonn potential that has become known as the CDBonn99.

This work was supported in part by the SFB 382 of the Deutsche 
Forschungsgemeinschafti,  the U.S.\ National Science
Foundation under Grant No.~PHY-9603097, the DGICYT (Spain) Grant PB95-1249 and the 
Program GRQ96-43 from Generalitat de Catalunya.

\begin{table}
\begin{tabular}{c|rrrrr}
& $E_{BHF}$ & $E_{HF}$ & $\Delta E_{ISB}$ & $E_{Sym}$ & $E_{Sym}^{np}$ \\
\hline
CDBonn & -16.78 & 6.06 & 0.329 & 30.65 & 30.45 \\
CDBo99 & -16.83 & 6.09 & 0.370 & 30.56 & 30.36 \\
ArgV18 & -15.58 & 31.73 & 0.282 & 29.74 & 29.77 \\
Nijm1  & -15.48 & 13.04 & 0.329 & 29.43 & 29.16 \\
Nijm2  & -13.71 & 35.62 & 0.221 & 28.08 & 27.84 \\
Reid93 & -14.28 & 35.46 & 0.218 & 28.24 & 28.21 \\
\end{tabular}
\caption{Energies calculated for nuclear matter with Fermi momentum $k_F$ = 1.36
fm$^{-1}$. Results are listed for the energy per nucleon calculated in BHF
($E_{BHF}$) and Hartree-Fock ($E_{HF}$) approximation, the loss of energy per
nucleon due to the breaking of isospin-symmetry [$\Delta E_{ISB}$ as defined in
Eq.~(\protect{\ref{eq:one}})], the symmetry
energy with ($E_{Sym}$) and without ($E_{Sym}^{np}$) inclusion of ISB.
All entries are in MeV.}
\label{tab:one}
\end{table}

\vfil\eject

\vfil

\end{document}